\documentclass[preprint,review,3p]{elsarticle}
\usepackage{graphicx}% Include figure files
\usepackage{dcolumn}% Align table columns on decimal point
\usepackage{bm}% bold math
\usepackage[english]{babel}
\usepackage{upgreek}
\usepackage{amsmath}
\usepackage{amssymb}
\usepackage{color,soul}
\usepackage{latexsym}
\usepackage{hyperref}
\hypersetup{
     colorlinks   = true,
     linkcolor    = blue,
}
\usepackage[utf8]{inputenc}
\usepackage[nottoc]{tocbibind}

\journal{Physica A}

\begin{document}

\begin{frontmatter}

\title{
Shape-velocity correlation defines polarization \\ in migrating cell simulations
}

\author[1]{Gilberto L. Thomas\corref{cor1}}
\ead{glt@if.ufrgs.br}

\author[1]{Ismael Fortuna}

\author[1]{Gabriel C. Perrone}

\author[2]{Fran\c{c}ois Graner}

\author[1,3,4]{ \\ and Rita M.C. de Almeida\corref{cor1}}
\ead{rita@if.ufrgs.br}

\address[1]{Instituto de F\'{\i}sica, Universidade Federal do Rio Grande do Sul, Porto Alegre, RS, Brazil} 
\address[2]{Mati\`ere et Syst\`emes Complexes, 
UMR7057 CNRS and Universit\'e de Paris-Diderot}

\address[3]{Instituto Nacional de Ci\^encia e Tecnologia: Sistemas Complexos, \\ Universidade Federal do Rio Grande do Sul, Porto Alegre, RS, Brazil} 

\address[4]{Programa de P\'os-Gradua\c{c}\~ao em Bioinform\'atica, Instituto Metr\'opole Digital, \\ Universidade Federal do Rio Grande do Norte, Natal, RN, Brazil}

\cortext[cor1]{Corresponding author}

\begin{abstract}
Cell migration plays essential roles in development, wound healing, diseases, and in the maintenance of a complex body. Experiments in collective cell migration generally measure quantities such as cell displacement and velocity. 
The observed short-time diffusion regime for mean square displacement in single-cell migration experiments on flat surfaces calls into question the definition of cell velocity and the measurement protocol. Theoretical results in stochastic modeling for single-cell migration have shown that this fast diffusive regime is explained by a  white noise acting on displacement on the direction perpendicular to the migrating  cell polarization axis (not on velocity). The prediction is that only the component of velocity parallel to the polarization axis is  a well-defined quantity, with a robust measurement protocol.
Here, we ask whether we can find a definition of a migrating-cell polarization that is able to predict the cell's subsequent displacement, based on measurements of its shape.
Supported by experimental evidence that cell nucleus lags behind the cell center of mass in a migrating cell, we propose a robust parametrization for cell migration where the distance between cell nucleus and the cell's center of mass defines cell shape polarization.  We tested the proposed methods by applying to a simulation model for three-dimensional cells performed in the CompuCell3D environment, previously shown to reproduce biological cells kinematics migrating on a flat surface. 
\end{abstract}


\end{frontmatter}

\section{\label{sec:Intro}Introduction}

% Motivation 
Polarization plays a key role in cell migration, both regarding single-cell movement on a flat substrate, or cells collectively moving in a tissue. In fact, polarization is important in every active matter model \cite{TenHagen2011,Romanczuk2012}, since the particle's activity necessarily breaks spatial symmetry. Although there are many different polarization definitions, as the one relative to the basal-to-apical axis in epithelia, in this paper we focus on cell polarization related to the rear-to-front axis associated to eukaryotic cell migration. Here we pay special attention to its correlation to cell velocity. In what follows, we carefully define what we mean by cell polarization and its measurement procedure, as well as how we define and measure cell velocity from a succession of snapshots of cell position obtained from  numerical simulations. Our proposition for polarization measurement procedure should serve to infer cell velocity modulus and direction in both simulations and experiments. 

Migrating cell polarization has its roots in the biological organization inside the cell and hence can be taken as a possible indicator of the cell movement. In fact, Maiuri 
and collaborators showed that there is a Universal Coupling between Speed and Polarization (UCSP) \cite{Maiuri2015}, such that both polarization and speed are correlated. UCSP is based on the fact that cell speed and persistence depends on a general advection of polarity ingredients, such as myosins or other molecules, with retrograde flow of actin in respect to the cell's center of mass (for a review, see references \cite{Insall2009,Ridley2011}). In this case, there must be a link between cell polarization and  kinematic measurements such as MSD curves or velocity auto correlation functions (VACF). To find this link, it is necessary to propose a robust cell speed measurement and provide a precise definition for cell polarization, that could be used in both single-cell and collective migration as a quantitative predictor for cell movement. Here we aim at defining 
a migration-related polarization  that could serve as a predictor for cell migration.

Velocity definition requires care.
Reports on quantitative experiments on single-cell migration on a flat substrate date back to 1920 when F\"urth \cite{Furth20} showed that the mean square displacement (MSD), represented by $\langle |\Delta \vec{r} |^2\rangle$, is obtained from the stationary  solution of an isotropic Ornstein-Uhlenbeck problem for cell velocity \cite{Uhl1930} now known as F\"urth equation:
\begin{equation}
\label{Eq:Furth}\langle |\Delta \vec{r} |^2 \rangle = 4D\left[ \Delta t -P\left( 1 -\exp\left(-\Delta t/P\right) \right) \right] \; ,
\end{equation}
where  $\Delta t $ stands for the time interval used to obtain the squared cell displacement $|\Delta \vec{r} |^2$. The symbols $ \langle \; \; \rangle $ represent averages over different times and trajectories, after the stationary state is reached. $D$ stands for the diffusion coefficient of the cell, reflecting the fact that for long $\Delta t $ we have $\langle |\Delta \vec{r} |^2 \rangle\sim 4D \Delta t $. This equation represents the sum over the displacements measured in two equivalent, perpendicular directions. On the short-time interval limit, Eq. \ref{Eq:Furth} yields a ballistic regime for which $\langle |\Delta \vec{r} |^2\rangle\sim \frac{2D}{P} \Delta t^2 $. Here, $P$ stands for the time scale associated to the transition from  ballistic to  diffusive regime, and is known as the persistence time of the model. This equation is obtained from the stationary, exact solution for a stochastic, 2D, isotropic Langevin problem for cell velocity (when the memory on the initial state has been lost), that may be written as 
\begin{equation}
\label{Eq:Langevin}
\frac{d\vec{v}}{dt}=-\frac{\gamma}{m} \vec{v} + \vec{\eta}(t),
\end{equation}
where $\gamma$ represents the friction of a particle moving in a fluid, $m$ the particle mass and $\vec{\eta}(t)$ is a white, 2D-vector noise, whose components have a standard deviation given by  parameter $g$, defined by $\langle \eta_i (t) \eta_j (t\prime) \rangle  = g \delta_{ij} \delta (t-t\prime)$. Here, $P=\frac{m}{\gamma}$ and $D=\frac{g}{2m\gamma}$. 
Observe that this equation is isotropic in space. Hence, it should not come as a surprise that
a direct consequence of the anisotropy of migrating cells comes in the form of a deviation in the MSD equation observed in both experiments \cite{Thomas2019} and simulations \cite{Fortuna2019}, which follows instead the modified F\"urth equation:
\begin{equation}
\label{Eq:ModFurth}
\langle |\Delta \vec{r} |^2\rangle= 2D \left[ \Delta t -P\left( 1 -\exp\left(-\Delta t/P\right) \right) \right]+ \frac{2DS}{1-S} \Delta t\; ,
\end{equation}
where $0 \le S < 1$ is a non-dimensional parameter. The modified F\"urth equation presents a further diffusive regime for short time intervals, since for short timescales $\langle |\Delta \vec{r} |^2\rangle\sim \frac{2DS}{1-S} \Delta t $,  indicating an additional, fast diffusive regime.  $SP$ gives the time scale for which the systems present a transition from this fast, short time interval diffusive regime to a ballistic-like regime, with $P$ signaling a second regime transition, from the ballistic-like to a slow, long time interval diffusion. By defining natural units for time ($P$) and length ($\sqrt{2DP/(1-S)}$), Eq. \ref{Eq:ModFurth} may be recast in terms of non-dimensional quantities $\langle |\Delta \vec{\rho}|^2 \rangle=\langle |\Delta \vec{r} |^2\rangle (1-S)/2DP$ and $\Delta \uptau= \Delta t/P$ as follows
\begin{equation}
\label{Eq:ModFurth2}
    \langle |\Delta \vec{\rho}|^2 \rangle = \Delta \uptau -(1-S) \left[ 1-\exp(-\Delta \uptau) \right] .
\end{equation}

The short-time interval diffusion implies that instantaneous velocity is not a well-defined quantity \cite{Thomas2019}, and the Langevin problem, written in the form of a time derivative for particle velocity, cannot  yield the modified F\"urth equation, Eq. \ref{Eq:ModFurth2}. 
 De Almeida {\it et al.} \cite{deAlmeida2020} proposed a set of equations where in a given instantaneous direction  (that we shall call polarization direction) 
 the dynamics is equivalent to a one dimensional Langevin problem, while in the perpendicular direction(s) the dynamics is driven by a Gaussian noise in the equation for displacement (not velocity), that is, a translational noise. A dynamic equation for the angle describing the direction of the Langevin dynamics completes the model. In Ref. \cite{deAlmeida2020} the authors analytically obtain the asymptotic solution and show that the modified F\"urth equation is the exact form for the MSD. In this model, the short-time diffusive motion comes from the translational noise in the direction perpendicular to polarization, that is, the deviation from the pure F\"urth MSD behavior observed in both simulations and experiments is a direct consequence of the spatial-symmetry break in the kinematics of migrating cells.
 
The consequence of a diffusive motion at short timescale is that as $\lim_{\Delta t\rightarrow 0} \frac{\Delta \vec{r}}{\Delta t}$ \ does not converge \cite{deAlmeida2020}: an experimentalist may get different values for $\frac{\Delta \vec{r}}{\Delta t}$, depending on the value of $\Delta t$ chosen to measure $\Delta \vec{r}$, regardless how small $\Delta t$ is.

% Approach and Outline

In a recent paper, Fortuna and collaborators \cite{Fortuna2019} proposed a simulation model for a three dimensional cell migrating on a flat substrate, using the 
cellular Potts model (CPM) 
\cite{Hogeweg2002,Graner1992,Glazier1993} 
in which, like in an experimental image, each cell is represented by a connected set of pixels.
This model has successfully reproduced the migration kinematics observed in experiments \cite{Thomas2019}. The computational CC3D simulation project, along with the instructions to run it, can be downloaded from \cite{Code2021}. Here we propose using this model to analyze different geometrical definitions of cell polarization as related to migration. We validate the proposed definitions by their correlation to the cell mobility. Next section defines quantities and their measurement procedures, section III presents the results, and in section IV we discuss and present our conclusions.

\section{Definitions and methods}

\subsection{\label{sec:Simu}Simulation model}

We use the cellular Potts model implemented in the open source CompuCell3D (CC3D) simulation environment \cite{Swat2012}.
We build a cell as composed of three compartments, each being a connected cluster of sites with an identifying label $\sigma$, that is related to the cell (the medium and substrate are  treated by the algorithm as special types of cells \cite{Fortuna2019}). This feature makes the CPM unique when trying to simulate detailed cell shape contours and fluctuations. A second label, $C$, is used to  identify the cell compartment: nucleus ($C=1$), cytoplasm ($C=2$), or lamellipodium ($C=3$). In this an energy-like function is written as
\begin{equation}
\label{Eq:Energy}
E=E_{interface}+E_{volume} \; \;
\end{equation}
where 
\begin{equation}
\label{Eq:EInterface}
E_{interface}=\sum_{\vec{r}}\sum_{\vec{s}\left(\vec{r}\right)} 
J\left( \vec{r}, \vec{s}\right),
\end{equation}
where $J\left( \vec{r}, \vec{s}\right)$ depends on the site labels $\sigma$ and $C$, and
is the interface energy per lattice-site surface between neighboring lattice sites at sites at  $\vec{r}$ and $\vec{s}$. The sum over $\vec{s}(\vec{r})$, runs over the 4th-neighbor range around $\vec{r}$ (32 neighbors) to reduce lattice anisotropy \cite{Holm1991}. $J=0$ is assumed for neighboring lattice sites that belong to the same cell and compartment. In all other cases, we set interface energies as positive (i.e. the $J$s are ferromagnetic).
We  choose  $J$ values to  ensure that the cytoplasm always surrounds the nucleus and the lamellipodium remains attached to the cytoplasm, substrate and medium.
The volume term  on the right hand side (r.h.s.) of Eq. \ref{Eq:Energy} constrains the volumes of the cell compartments to be close to their reference volumes:
\begin{equation}
\label{Eq:EVolume}
E_{target\ volume}=\sum_{C=1}^{3}{\lambda_C\ {\big(V_C-V_C^{target}\big)}^2},
\end{equation}
where $V_C$ is the current volume of the $C$-th cell compartment, $V_C^{target}$ is its target volume, and $\lambda_C$ is the inverse compressibility of the compartment. 

Configuration changes are performed as usual in Monte Carlo simulations: after randomly picking a pair of nearest neighbors cell-lattice sites, we tentatively 
copy labels $\sigma$ and $C$ from the first to the second lattice site and calculate the  change in energy $\Delta E$.   If $\Delta E \leq 0$, we accept the change, and if $\Delta E > 0$, we accept the change with probability $\mathrm{exp}(-\Delta E/T_m)$, where $T_m$ is a parameter describing the amplitude of cell-membrane fluctuations. A Monte Carlo Step (MCS) consists of a number of  copy attempts equal to the number of sites in the lattice.

\begin{figure}[h]  
\centering
\includegraphics[width=.5\textwidth]{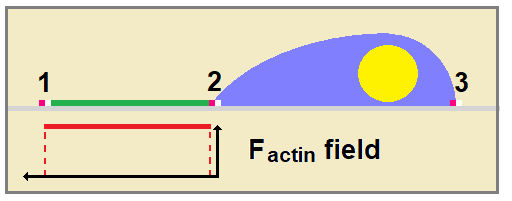}
\caption{\label{fig:Fig1NEW} {\bf Crawling dynamics in an {\it in silico} cell. }The cell compartments are: lamellipodium in green, cytoplasm in blue, and nucleus in yellow. Three steps create a polarization axis that favors migration leftwards: 1 - The $F_{actin}$ field, which effectively non-zero at lamellipodium  lattice sites (red curve), favors lamellipodium lattice sites to be copied over medium lattice sites. This increases lamellipodium compartment volume and extends it to the left. 2 - Since there is a volume constraint in the lamellipodium compartment, there is now a preferential copy of cytoplasm lattice sites over lamellipodium ones, which  decreases lamellipodium compartment
volume and increases cytoplasm compartment volume. 3 - Again there is a volume constraint on the cytoplasm compartment, which favors medium lattice sites to be copied over cytoplasm compartment ones. 
} 
\end{figure}

 Non-conservatives forces  require some modification to the usual CPM dynamics. We add a further term when calculating $\Delta E$, to extend the usual CPM 
 effective energy to describe the protrusive forces that F-actin polymerization exerts on the leading edge of the cell. After choosing a site $\vec{r}$ and its neighboring copy target site $\vec{s}$, we add the following term to $\Delta E$:
 \begin{equation}
\label{Eq:EFactin}
\Delta E_{F-actin}=\lambda_{F-actin}\left[F(\vec{s})-F(\vec{r})\right], 
\end{equation}
where the sum is taken only on the boundary between lattice sites lying in the lamellipodium compartment and the surrounding medium. By changing $\Delta E$ and consequently changing the probability that a lamellipodium site is copied over a medium site, Eq. \ref{Eq:EFactin} simulates a force to the membrane in the direction of the copy attempt (for more details, see Ref.\cite{Fortuna2019}).  $F\left(\vec{r}\right)$ is a field specifying the concentration of  F-actin, and $\lambda_{F-actin}>0$ is the force per area per unit F-actin field. The actin concentration field obeys the simplified reaction-diffusion equation, as described in Ref.\cite{Fortuna2019}, that is,
\begin{equation}
\frac{\partial F(\vec r,t)}{\partial t} = D_F \nabla^2 F(\vec r,t) + k_s\delta(C(\vec{r},t)-3) - k_dF(\vec r,t),
\end{equation}
where $D_F$ is an effective diffusion constant, $k_{s(d)}$ represents the polymerization (depolymerization) rate of F-actin fibers, and the $\delta$-function says that only lamellipodium creates them. This equation is solved during the simulations by a tool included in the CC3D environment. Details on CC3D PDE solvers can be read in \cite{PDEsolver}.  The final form assumed by the F-actin field is illustrated in Fig.\ref{fig:Fig1NEW}: the parameters choice ensures that F-actin field is nearly constant on lamellipodium sites and zero otherwise.

The term for $\Delta E_{F-actin}$ together with Eq. \ref{Eq:Energy} i) increase the rate at which lamellipodium  is copied over cytoplasm; ii) with the lamellipodium volume increase, the copy of cytoplasm sites over lamellipodium is favored (due to lamellipodium target volume term), and iii) the same happening to the rate with medium lattice sites overwrite cytoplasm lattice sites, creating a polarization axis that drives cell migration in the direction of the lamellipodium compartment. This polarization and its  consequences simulate the effect of substrate forces and the interaction with extra-cellular matrix through adherent complexes in real cells, in a qualitative and simplified way. For a more detailed model  that  considers elastic substrates and the explicit role of actomyosin fibers and adhesion complexes, we refer the reader to the works by Zhong and collaborators \cite{Zhong2013,Zhong2014}  and He and collaborators \cite{He2014} .

\begin{figure}[ht]  
\centering
\includegraphics[width=.9\textwidth]{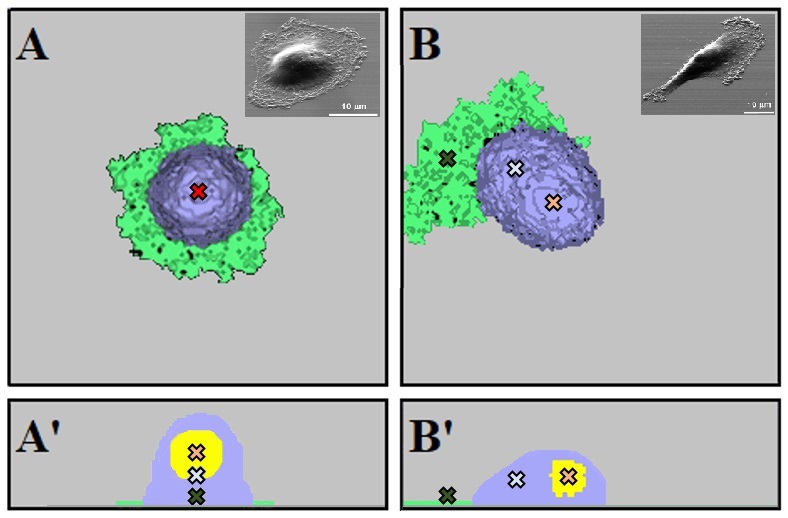}
\caption{\label{fig:Fig2NEW} {\bf Shape polarization definitions. }
Snapshots of simulated cells.  A (view from above) and A' (vertical cross section) represent a cell whose spatial symmetry has not been broken.  The centers of mass of nucleus (yellow), cytoplasm (blue), and lamellipodium (green) almost coincide (red cross). In A', orange, light gray, and green crosses locate the centers of mass of, respectively, nucleus, cytoplasm, and lamellipodium. B (view from above) and B' (vertical cross section) represent a cell with the broken spatial symmetry.  In B, the projection on the substrate plane of the centers of mass do not coincide, and Nucleus lags behind. The centers of mass of these three compartments can be combined to produce different definitions of polarization vectors.  In the insets, for comparison, experimental images from \cite{Chen2015}.
} 
\end{figure}

Figure \ref{fig:Fig1NEW} illustrates the crawling dynamics in an {\it in silico} cell, and Figure \ref{fig:Fig2NEW} shows two snapshots of them, as created by the CC3D, migrating cells (from Ref.\cite{Chen2015} ), depicting different morphologies that they may assume. Observe that lamellipodium may be engulfing the cytoplasm or concentrated at one side of the cell. In the first case, lamellipodium tends to grow over medium in all directions, producing a short timescale diffusive regime. In the second situation, lamellipodium has a preferred direction to grow over medium sites and creates the polarization axis illustrated in \ref{fig:Fig1NEW} that induces a preferred motion direction, contributing to a non-diffusive motion.
 
 Larger values of $\lambda_{F-actin}$ favors localizing lamellipodium at one side of the cytoplasm, due to a local-excitation, global-inhibition loop: the more lamellipodium at one side of the cytoplasm, the more lamellipodium grows there (local excitation) and when more lamellipodium is created, the target volume term inhibits the creation of new lamellipodium sites everywhere (global inhibition). Consequently, larger values of $\lambda_{F-actin}$ favors polarization, decreasing the time interval where it is possible to detect the short time diffusive regime, scaled by $SP$.

\subsection{\label{sec:Polar} Polarization Definitions and Measurements}

As we show in this paper, the intuitive idea that front-rear cell polarization should be able to predict cell movement critically depends on how we define this movement. Displacement and instantaneous velocity are two possibilities that are interchangeable when velocity is a well defined vector quantity, which frequently is not the case for migrating cells. Instantaneous velocity is obtained by taking the limit of the displacement per unit time interval $\Delta t$ as $\Delta t \to 0$. This limit does not exist when the short time interval diffusive term is present (see Eq. \ref{Eq:ModFurth}). In discussing the polarization correlation with cell movement we must then use a mean velocity, where the time interval is kept at a finite value. Because of that, in what follows, we use the terms  displacement, velocity and mean velocity meaning different quantities, that cannot be used to characterize the same measurement. Also, we analyze both direction and modulus of mean velocity correlations with polarization. We use the term mean speed for mean velocity modulus, keeping the term velocity for the corresponding vector quantity.

To define a migration-related polarization that could serve as a predictor for cell migration, the choice of the candidate quantities to be further analyzed  is based on two biological observations: {\it i}) there is a universal coupling between speed and polarization as described by Maiuri and collaborators \cite{Maiuri2015}, that was later applied to explain the retrograde actin flow in migrating cells \cite{CallanJones2016}; and {\it ii}) the nucleus is positioned at the rear of a migrating  cell \cite{Gundersen2018}. Both observations are explained by the  cytoskeleton  arrangement in migration, where  acto-myosin fibers and microtubules help in both dragging the cell nucleus and in  the trafficking of the necessary proteins, creating the required protein gradients that support the trailing edge and rear contraction, responsible for cell migration. 

Here we consider the following possible measures for the cell polarization $\vec{\Pi}$, that explicitly consider the nucleus, to grasp the experimental observation by Gundersen and Woman \cite{Gundersen2018} that in migrating cells the nucleus lags behind:
\begin{eqnarray}
\label{Eq:PolDef}
\vec \Pi_{L-N}(t) &=& \Big( \vec r_L(t)-\vec r_N(t)\Big)_{xy},\nonumber \\
\vec \Pi_{L-CN}(t) &=& \Big( \vec r_L(t)-\vec r_{CN}(t) \Big)_{xy}, \nonumber \\
\vec \Pi_{C-N}(t) &=&  \Big( \vec r_{C}(t)-\vec r_N(t) \Big)_{xy}, \\
\vec \Pi_{CN-N}(t) &=&  \Big( \vec r_{CN}(t)-\vec r_N(t) \Big)_{xy}, \nonumber \\
\vec \Pi_{C-CN}(t) &=&  \Big( \vec r_{C}(t)-\vec r_{CN}(t) \Big)_{xy}, \nonumber
\end{eqnarray}
where the indices $C$, $L$, and $N$ refer to the different  cell compartments (Cytoplasm, Lamellipodium, and Nucleus) and $CN$ refers to the combined compartment comprising Cytoplasm and Nucleus (Fig.~\ref{fig:Fig2NEW}), and $\vec r_{\beta}(t)$ for $\beta=L,C,N,CN$ refers to the center of mass of these compartments. 
Polarization vectors $\vec{\Pi}_\beta$, for $\beta=L-N$, $L-CN$, $C-N$, $CN-N$, and $C-CN$  are defined as the difference vector between the centers of mass of two compartments, projected on the substrate, as indicated by the subindices in the right hand side of Eqs. \ref{Eq:PolDef}.  At each MCS these quantities are recorded. We recall that by knowing $C$ and $N$, one can directly reconstruct  the combined compartment $CN$  comprising both Cytoplasm and Nucleus, and for instance:
\begin{eqnarray}
\vec \Pi_{C-N}(t) &=& \left ( \frac{V_C+V_N}{V_C}\right ) \vec \Pi_{CN-N}(t) \mbox{      and} \nonumber \\
\vec \Pi_{C-CN}(t) &=& \left ( \frac{V_N}{V_C}\right ) \vec \Pi_{CN-N}(t),
\end{eqnarray}
where  $V_C$ and $V_N$ are the compartment volumes. From the above equations, we have that\linebreak $\vec \Pi_{C-CN}(t) = \vec \Pi_{C-N}(t) - \vec \Pi_{CN-N}(t)$. Consequently, $\vec \Pi_{C-N}$, $\vec \Pi_{CN-N}$ and $\vec \Pi_{C-CN}$ contain the same information. However, in experiments we may easily assess  position of the center of mass of the combination of  cytoplasm and  nucleus by the cell's geometrical center. In what follows we present results considering four definitions for cell polarization, disregarding $\vec\Pi_{C-CN}$.

To estimate the performance of the polarization measures as a proxy to cell displacement, we must estimate the correlation between these vectors and the subsequent cell displacement. However, we must first choose a suitable time interval. For too small  time intervals (smaller than $SP$), displacements may be dominated by random fluctuations in the direction perpendicular to cell polarization \cite{Thomas2019,Fortuna2019,deAlmeida2020}. For too large  time intervals, larger that the persistence time $P$, displacement may also decorrelate with cell polarization. Hence, to evaluate the performance of cell polarization definitions as predictors of cell displacement, we must first choose an adequate time interval. We shall define as $\Delta \uptau_{opt}$ the time interval for obtaining cell displacement that has the highest correlation with cell polarization.

A theoretical estimate of  $\Delta \uptau_{opt}^{theor}$ is obtained by finding the time interval that maximizes the slope of the log-log curve $\langle |\Delta \vec{\rho}|^2 \rangle $, given in Eq. \ref{Eq:ModFurth2}. The slope $\upalpha$ is given by
\begin{equation}
\label{Eq:alpha}
\upalpha = \frac{d \log(\langle |\Delta \vec{\rho}|^2 \rangle)}{d \log(\Delta \uptau)} = \frac{\Delta\uptau(1-(1-S)e^{-\Delta\uptau})}{\Delta\uptau-(1-S)(1-e^{-\Delta\uptau})}\, .
\end{equation} 
 The inclination $\upalpha$ assumes values in the interval [1,2], passing through a maximum  for  $ S\leq \Delta \uptau \leq 1$, that is obtained by numerically solving a transcendental equation. We assume that the value of $\Delta \uptau$ that maximizes $\upalpha$  defines a suitable time interval to perform correlation measurements. To test this hypothesis,
 we must compare with adequate correlation measurements as follows.
 
 To determine the  time intervals for which the short-time random fluctuations dominate the dynamics, we monitor mean speed $u(\uptau, \delta)$, defined as the modulus of the mean velocity $\vec{u}(\uptau, \delta)$ for a finite time interval $\delta$ as
\begin{equation}
\label{Eq:u}
  \vec u(\uptau, \delta) =\frac{\vec{\rho}(\uptau + \delta) -\vec{\rho}(\uptau)}{\delta}\, ,
\end{equation}
where $\vec\rho$ is the cell position in natural units, and the average of both mean velocity $\langle \vec{u}\rangle_{\delta}$  and mean speed $ \langle  u \rangle_{\delta}$ are taken over all time points $\uptau$ and different runs, after the stationary state has been reached. Since for small $\delta$ we expect the random, short-time noise to dominate the dynamics, we also expect  $ \langle  u \rangle_{\delta}$ to diverge as $\delta \rightarrow 0$. On the other hand, when $\delta = \Delta \uptau_{opt}^{theor}$ we expect that the mean speed and polarization correlate.

At each MCS, we recorded the center of mass of each compartment and of the combined cytoplasm and nucleus compartment together with the different polarization definitions, as given by Eqs. \ref{Eq:PolDef}. From there it is possible to obtain the angle between polarization at a given time and the mean velocity in the subsequent time interval $\delta$.  We monitor the best correlation between the directions of cell polarization and cell mean velocity by monitoring  the  velocity direction - polarization  correlation index $\xi_{\beta}$ for each polarization ${\beta}$ defined in Eqs. \ref{Eq:PolDef}, defined as
\begin{equation}
    \label{Eq:xi}
    \xi_{\beta}=\frac{\langle \cos{\theta_{\beta}}\rangle }{\langle \sin{\theta_{\beta}}\rangle} \;\;\; ,
\end{equation}
where $\langle \cdot \rangle$ represents average over the whole trajectory (after the stationary state has been reached) and over different simulation runs.  Index ${\beta}$ stands for the labels $CN-N$, $L-CN$, and $L-N$ so that $\theta_{\beta}$ represents the angle between cell displacement and the ${\beta}$-th polarization direction ($C-N$ is colinear to $CN-N$ and is not explicitly taken into account here). Here $\xi_{\beta}$ depends on the time interval $\delta$ used to obtain cell mean velocity and its values range from $0$, when polarization and the mean velocity  $ \vec{u}(\uptau, \delta) $ are always orthogonal, to $\infty$, when  polarization and the mean velocity $\vec{u}(\uptau, \delta)$ are always parallel.

Besides correlation between the adequate measure for polarization and mean velocity modulus and direction, we also assessed the dispersion around the predicted value for both velocity modulus and direction. In this way, besides predicting the expected value for $ \langle  u \rangle_\delta$, we also give the uncertainty we should expect from that prediction. For the modulus, we propose a relative error, $\upvarepsilon$, given as
\begin{equation}
\label{Eq:error_speed}
\upvarepsilon= \frac{\sigma_{speed}}{\langle u \rangle_{\Delta \uptau_{opt}^{theor} }}  \;\;\; ,
\end{equation}
where $\sigma_{speed}$ is the standard deviation of $|\vec{u}(\uptau,\Delta\uptau_{opt}^{theo})|$ over the trajectory.
 
 We monitor the dispersion in direction 
 using the standard deviation of $\theta_\beta$, denoted $\sigma_{\theta_\beta}$, since the average $\langle \theta_\beta \rangle $ is close to 0 and hence cannot provide a scale for dispersion. Anyway,  angles have  a finite natural scale of $2\pi$. Table S1 in Supplementary Materials Online lists the main variables and parameters used in this work.

\section{Results}

We first verify whether the polarization definitions presented in Eqs. \ref{Eq:PolDef} may be used as predictors for cell velocity by analysing our simulation results.  In this paper, we keep fixed  the  parameters associated with energy and temperature  as in Ref.\cite{Fortuna2019} and  varied parameters related to actin-force intensity ($\lambda_{F-actin}= 150, 175,$ and $200 $) and the lamellipodium volume fraction  ($\phi=0.05, 0.10, 0.20,$ and $0.30$). For each set of parameters, we run 10 simulations with $10^5$ MCS. We also considered three different cell sizes, $R_{cell}=10,15,$ and $20$ measured in $(\mbox{lattice sites})^{1/3}$, to test for finite size effects. The behavior is qualitatively the same and in the main text we considered $R_{cell}=15$ while all results concerning $R_{cell}=10,$ and $20$, are shown in Supplementary Materials Online. Table S2 in Supplementary Material Online shows the parameter sets and fitting constants. For more details in the simulations, we refer the readers to Ref. \cite{Fortuna2019}.

The simulations start with a spherical cell with no lamellipodium engulfing  a cubic nucleus resting over the substrate. The dynamics of the model is such that the cytoplasm sites touching the substrate may transform into lamellipodium sites until the lamellipodium target volume is attained. Due to the energy parameters $J$ between sites belonging  to different compartments or medium and substrate, the lamellipodium spreads over the substrate, forming a thin layer  that surrounds cytoplasm, as shown in Fig. \ref{fig:Fig2NEW} A (see also Fig.4 from Ref. \cite{Fortuna2019}). For the parameters chosen for this manuscript, the lamellipodium eventually assumes a configuration where it localizes at one side of the cell, spontaneously breaking symmetry and initiating migration. The asymmetric lamellipodium configuration drives migration and its continuous direction changes is reponsible for the observed long time interval diffusion. Usually this behavior has been characterized by calculating the MSD. The transient from the initial configuration is detectable by the relaxation of the MSD curve to a stationary form, when the memory from the initial configuration is lost. Here, we have waited 1000 MCS before recording the data that we use in the analyses. After this simulation time MSD curves do not change any more.  For more details in the simulations, we refer the readers to Ref. \cite{Fortuna2019}.

\begin{figure}[h]
\centering
\includegraphics[width=.75\textwidth]{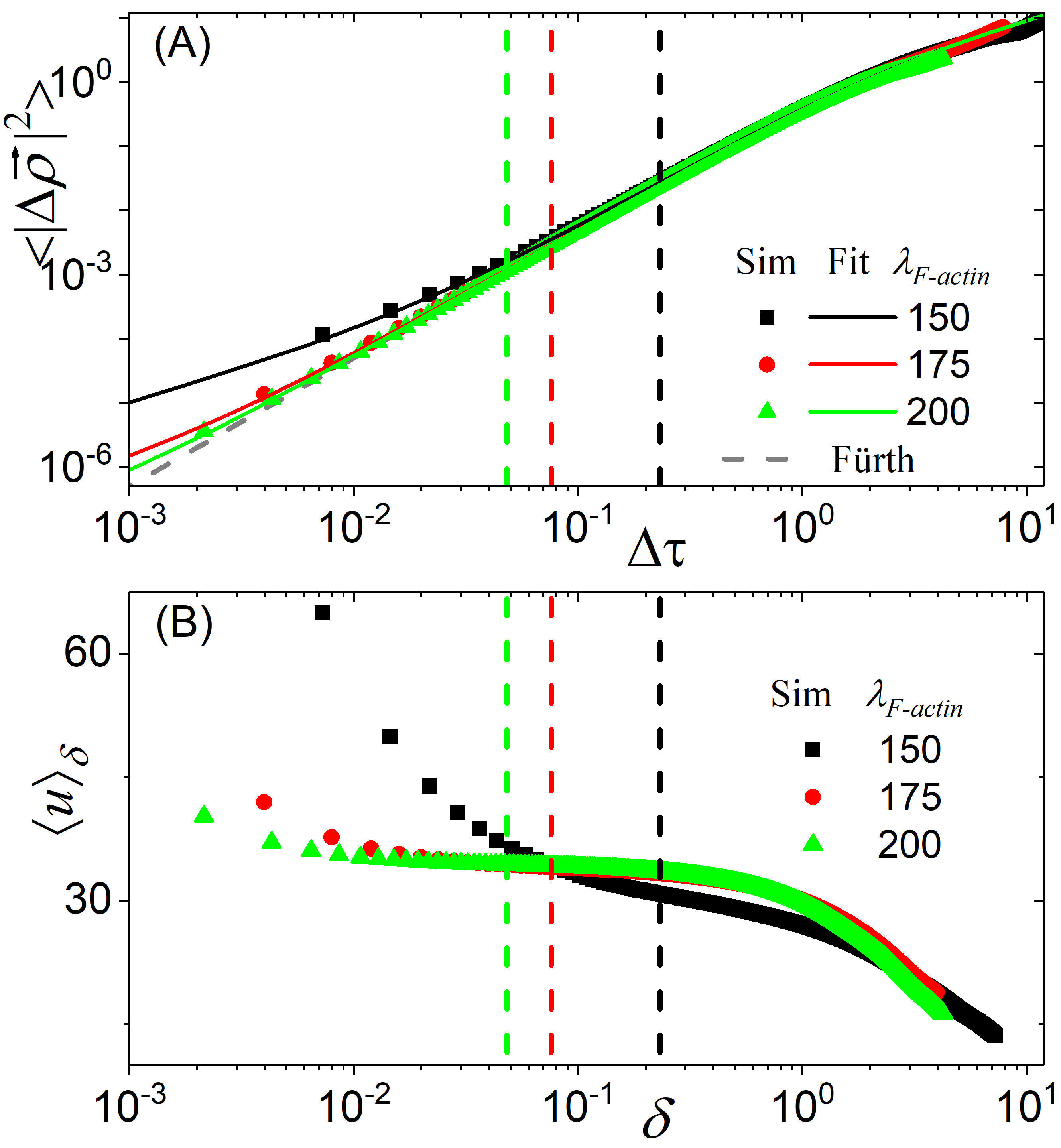}
\caption{\label{Fig:Fig3NEW} {\bf MSD and average mean speed curves reveal departure from classic F\"urth equation.}   (A)  $\langle |\Delta \vec{\rho}|^2 \rangle$  $versus$ time interval ($\Delta\uptau$) in comparison with F\"urth's equation (grey dashed line).   (B) $\langle u \rangle_\delta$ $versus$ $\delta$. The dots indicate the predicted value for the optimal time interval and maximum exponent for the MSD curves, corresponding to the values of $S$ obtained for the simulations. The curves in (A) and (B) stand for different parameter sets; each curve is averaged over 10 simulation samples for $R_{cell}=15$, $\phi=0.10$, for $\lambda_{F-actin} =150$ (black squares), $175$ (red circles), $200$ (green triangles). 
The dashed vertical lines indicate corresponding values for $\Delta \uptau_{opt}^{theor}$ in both panels. Standard errors in (A) and (B) are of the order of $10^{-2}$ or less  of the  respective values of $\langle |\Delta \vec{\rho}|^2 \rangle$ and $\langle u \rangle_\delta$ and are smaller than the symbols size.} 
\end{figure}

Figure \ref{Fig:Fig3NEW} shows the results for a selected set of parameters, as indicated.  In panel A, the mean square displacement $\langle |\Delta \vec{\rho}|^2 \rangle$ is plotted  $versus$ time interval ($\Delta\uptau$), both given in natural units, together with the fits using the modified F\"urth equation (Eq. \ref{Eq:ModFurth2}). The deviation from the original F\"urth equation (dashed, light gray line) for small time intervals is clearly visible. In panel B, we present the average mean speed $\langle u\rangle _\delta$ as a function of $\delta$. Observe that when $\delta$ is too small, mean speed diverges as expected, signaling a diffusive dynamics. In both panels A and B, the dashed vertical lines indicate the value of $\Delta \uptau_{opt}^{theor}$ for which the corresponding  $\langle |\Delta \vec{\rho} |^2 \rangle$ (in panel A) presents the maximum exponent $\upalpha_{opt}$, as predicted by maximizing Eq. \ref{Eq:alpha}. Values for $\Delta \uptau_{opt}^{theor}$ for all simulations are presented in Table S1 in Supplementary Materials Online, where we also present figures for all parameter sets similar to panels A and B (Figs. S1-S3).

\begin{figure}[h]
\centering
\includegraphics[width=.7\textwidth]{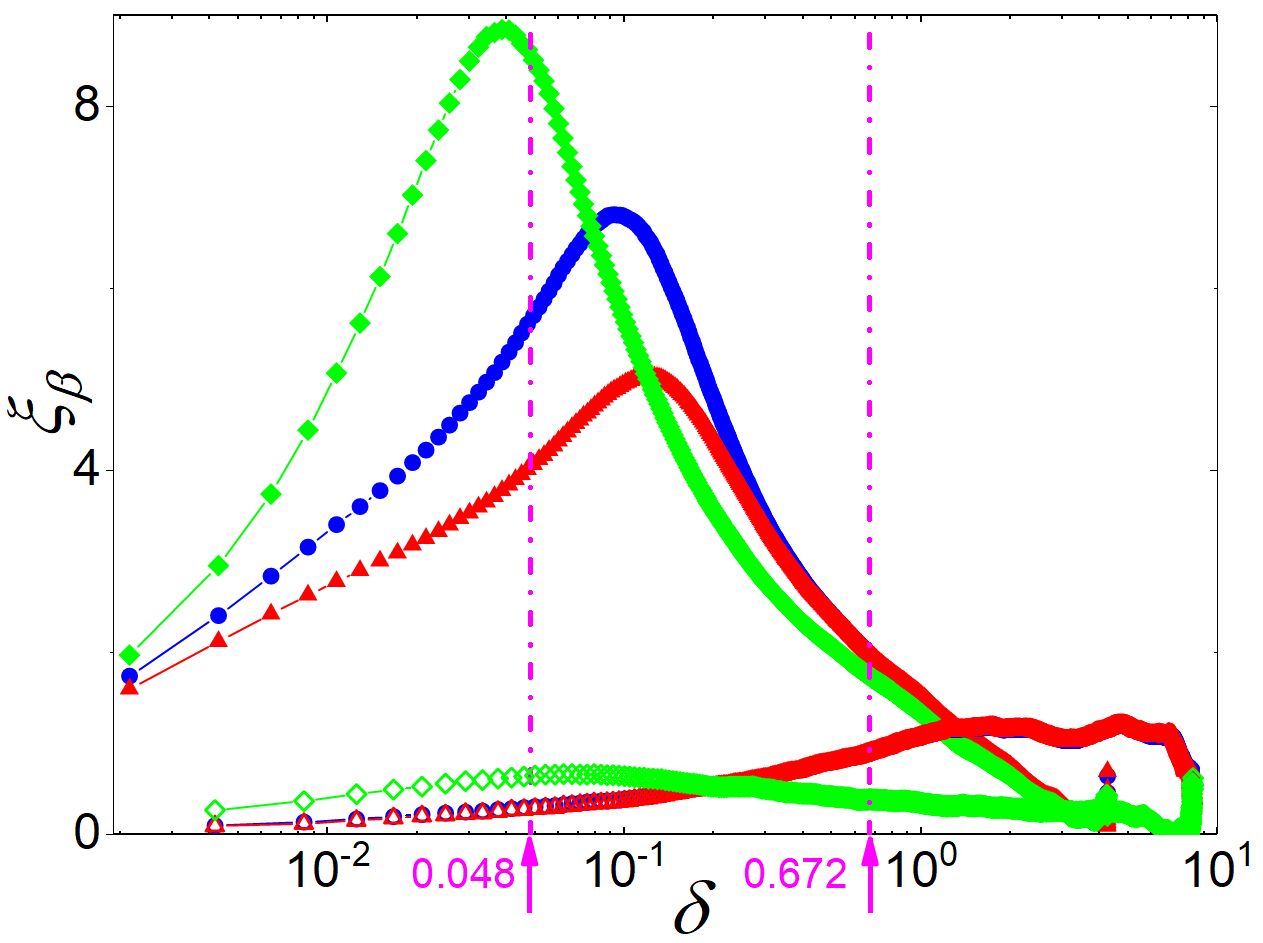}
\caption{\label{Fig:Fig4NEW} {\bf $\vec{\Pi}_{CN-N}$ is strongest correlator with mean velocity when calculated using $\delta = \Delta \uptau_{opt}^{theor}$.} $\xi_{\beta}$ as a function of the time interval $\delta$ used to obtain the displacement direction, considering the three polarization definitions: $CN - N$ (green diamonds), $L - CN$ (blue circles), and $L - N$ (red triangles), for two sets of parameters with $R_{cell}=15$: $\phi=0.10$, $\lambda_{F-actin} =200$ (filled symbols, $\Delta\uptau_{opt}^{theor} = 0.048$) and $\phi=0.30$, $\lambda_{F-actin} =150$ (open symbols, $\Delta\uptau_{opt}^{theor} = 0.672$). The pink arrows indicate the $\Delta\uptau_{opt}^{theor}$ for both parameter sets. Standard errors are smaller than the symbols size. Similar panels considering different values of $\phi$ for $R_{cell}=10,15,$ and $20$ are presented in Supplementary Materials Online.}
\end{figure}

Figure \ref{Fig:Fig4NEW} presents  $\xi_{\beta}$, defined in Eq. \ref{Eq:xi}, as a function of $\delta$, the time interval used to calculate cell displacement for three definitions of polarization and for two sets of simulations parameters. $\vec{\Pi}_{C-N}$ is not shown because its direction is identical to  $\vec{\Pi}_{CN-N}$ direction. What this figure shows us two important points, as follows.

First it shows that for $\phi= 0.10$  and  $\lambda_{F-actin}= 200$ the polarization definition whose velocity-polarization correlation index $\xi_{\beta}$ has its maximum more coincident with the theoretical prediction $\Delta\uptau_{opt}^{theor}$ (indicated by the pink arrows) corresponds to either $\vec \Pi_{C-N}(t)$ or $\vec \Pi_{CN-N}(t)$. 

Second, the case $\phi= 0.30$  and  $\lambda_{F-actin}= 150$ requires special care.
For this case  $S=9.46 \times 10^{-2}$, implying {\it i}) a small value for the maximum exponent $\upalpha$ of their MSD curves (see Fig. \ref{Fig:Fig3NEW} A),  {\it ii}) a short, almost nonexistent plateau for $\langle u \rangle_{\delta} $ (see Fig. \ref{Fig:Fig3NEW} B) , and {\it iii}) a short ballistic-like regime in the MSD curves, also shown in Fig. \ref{Fig:Fig3NEW} A.  These three observations suggest that the cell never reaches a proper ballistic phase. In fact, the weak intensity of $\lambda_{F-actin}$ and the large lamellipodium fraction reduces the probability of a symmetry break and the cell preferentially stays in a configuration as shown in Fig. \ref{fig:Fig2NEW} A. The persistence time is hence not large and the time intervals for the fast and slow diffusive regimes are similar. It is then not surprising that all three definitions of polarization fail to yield a velocity-polarization correlation index whose maximum is near the predicted value of $\Delta \uptau_{opt}^{theor}$. Figs. S1-S6 in Supplementary Materials Online present the same information in Figs. \ref{Fig:Fig3NEW} and  \ref{Fig:Fig4NEW} for all parameter sets.

Figure \ref{Fig:Fig5NEW}  shows $\upalpha_{opt}$ (purple) and $\Delta \uptau_{opt}^{theor}$ (magenta) as functions of $S$. We have used all fitted $S$-values from Ref.\cite{Fortuna2019} to numerically  obtain $\Delta \uptau_{opt}^{theor}$ that maximizes $\upalpha$ in Eq. \ref{Eq:alpha}. We fitted these points as a power law. Figure \ref{Fig:Fig5NEW} shows the fit as a  magenta solid line, with dots representing the numerical solutions for those values of $S$ corresponding to each simulation of the whole set, including $R_{cell}=10$ and $20$. In purple we show the equivalent for the values of $\upalpha_{opt}$. The numerically obtained  $\Delta\uptau_{opt}^{theor}$ are well fitted by the power law $\Delta\uptau_{opt}^{theor} = 1.932(\pm 8.6 \times 10^{-3})\; S^{\,0.455(\pm 2.0 \times 10^{-3})}$ with $R^2>0.999$, while the points associated to $\upalpha_{opt}$ are fitted by $\upalpha_{opt}=2.022(\pm 2.39 \times 10^{-3})-1.115(\pm 4.56 \times 10^{-3}) S^{\,0.397(\pm 3.90 \times 10^{-3})}$, with $R^2>0.999$. Observe that, as $S$ approaches zero,  $\Delta \uptau_{opt}^{theor}$ approaches zero and $\upalpha_{opt}$ approaches 2: when the short-time diffusive regime disappears, the optimal time interval goes to zero, as it should, such that the MSD curve presents a maximum exponent equal to 2.  Also, as $\Delta\uptau_{opt}^{theor}$ increases, $\upalpha_{opt}$ decreases: when the fast diffusive regime is long ($S$ is large), the  MSD curve has a short ballistic-like regime, such that the maximum value of $\upalpha_{opt}$ does not reach values near 2. In fact, for $S>0.3$ we have $\upalpha_{opt}<1.3$.

\begin{figure}[h]
\centering
\includegraphics[width=.75\textwidth]{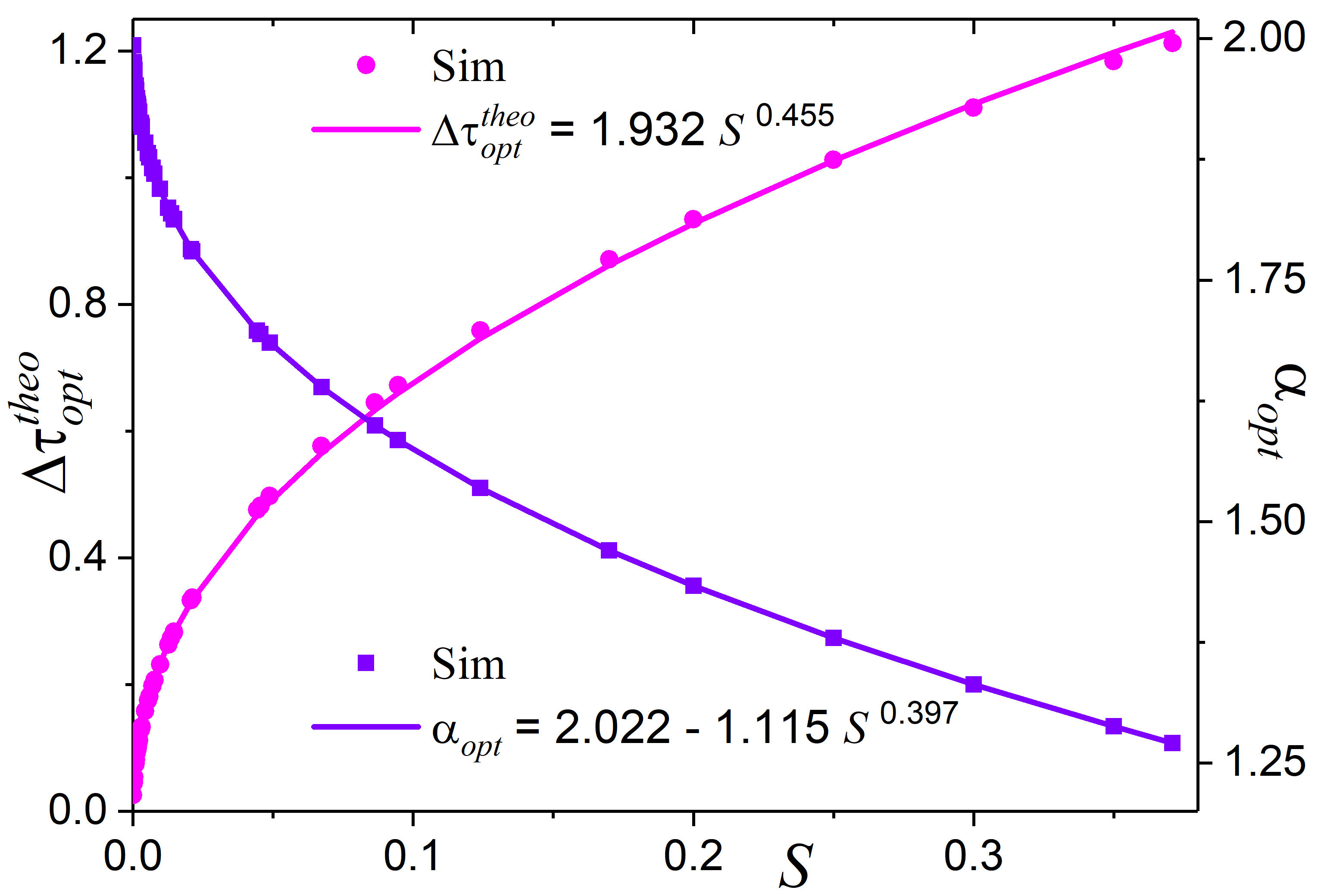}
\caption{\label{Fig:Fig5NEW}  Dots represent $\Delta\uptau_{opt}^{theor}$  and $\upalpha_{opt}$  as functions of $S$, as numerically obtained from the modified F\"urth equation, by maximizing Eq. 12 and using the values of $S$ obtained from the simulated MSD curves. Solid lines represent the fitting functions in the legend.}
\end{figure}

The consequence of the behavior shown in Figs. \ref{Fig:Fig3NEW} to  \ref{Fig:Fig5NEW} is that by fitting the MSD curve from simulations or experiments using the Modified F\"urth equation, it is possible to obtain $S$ and from there to obtain the time interval $\Delta\uptau_{opt}^{theor}$ for which the correlation between polarization direction and displacement direction is optimized.  
Based on the results shown in Fig. \ref{Fig:Fig4NEW},
we select $\vec{\Pi}_{CN-N}$ to proceed with our analyses. 

\begin{figure}[h]
\centering
\includegraphics[width=1\textwidth]{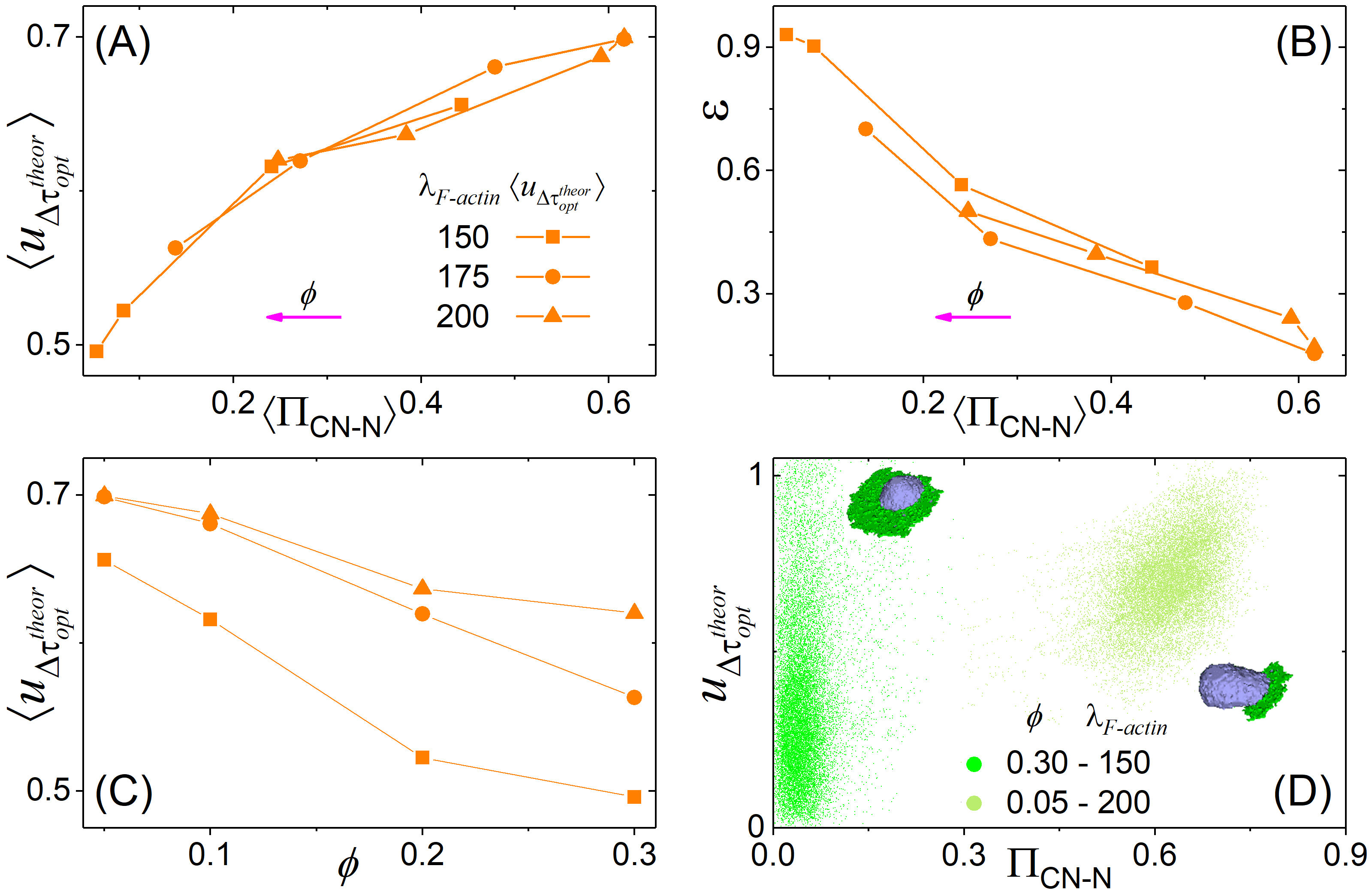}
\caption{ \label{fig:Fig6NEW} {\bf Performance of $\vec{\Pi}_{CN-N}$ as a predictor for cell speed.} (A) $\langle u_{\Delta \uptau_{opt}^{theor}}\rangle$  $versus$   $\langle {\Pi}_{CN-N}\rangle$ in units of cell radius  for the CN-N polarization definition Eq.(\ref{Eq:PolDef}).  (B) Relative error for the speed $\upvarepsilon$, as defined in Eq.(\ref{Eq:error_speed}), $versus$  $\langle {\Pi}_{CN-N}\rangle$. The pink arrows in (A) and (B) indicate the direction of increasing $\phi$. (C) The average speed $\langle u_{\Delta \uptau_{opt}^{theor}}\rangle$ $versus$ $\phi$. Each curve in (A,B,C) has 4 points, one for each value of $\phi$. All curves are for all parameter sets with $R_{cell}=15$. The legend in (A) applies to (B) and (C) as well. (D) Individual cell speed $u_{\Delta \uptau_{opt}^{theor}}$ $versus$ its polarization modulus $\Pi_{CN-N}$ measured for all points of the stationary trajectory and 10 different simulations, considering two parameter sets (representative simulated cells are shown). For $R_{cell}=10$ and $R_{cell}=20$ see Fig. S7 in Supplementary Materials Online.}
\end{figure}

Figure \ref{fig:Fig6NEW}  A shows the expectation value for mean speed $\langle u(\Delta  \uptau_{opt}^{theor}) \rangle $ and Fig.\ref{fig:Fig6NEW} B shows its relative error $ \upvarepsilon
$ $versus$ the average polarization modulus $\langle |\vec{\Pi}_{CN-N}| \rangle$ for 1980 points of 10 typical trajectories (19800 points), after the stationary state is reached. Each set of points is obtained using the same $\lambda_{F-actin}$ are linked with lines. In these lines, each point corresponds to a different lamellipodium fraction $\phi$ (0.05, 0.1, 0.2, and 0.3) increasing from right to left. 
This figure shows that  $\langle u(\Delta  \uptau_{opt}^{theor}) \rangle  $ increases with average polarization modulus, while the relative error in speed decreases: when $\langle |\vec{\Pi}_{CN-N}| \rangle$ is large, the cell migrates with higher speed and with less dispersion. 

Figure  \ref{fig:Fig6NEW} C shows the effect that increasing $\phi$ decreases the overall average of speed: as the cells have more lamellipodium, the symmetry break that allows the lamellipodium to efficiently drive cell migration is less frequent.

Figure  \ref{fig:Fig6NEW} D shows all measured points for two different parameter sets: $(\phi=0.3, \lambda_{F-actin}=150)$  that shows a very small polarization and $(\phi=0.05, \lambda_{F-actin}=200)$, where the cells are highly polarized most of the time. The cloud of points for each case illustrates that increasing $\langle |\vec{\Pi}_{CN-N}| \rangle$ correlates with increasing  $u(\Delta  \uptau_{opt}^{theor})$ and decreasing dispersion for more polarized cells. Figure  \ref{fig:Fig6NEW} D also presents a picture of a cell in a typical configuration from the simulations  for the two sets of parameters, evincing the different lamellipodium configurations for each case. 
 The clouds presented in Fig. \ref{fig:Fig6NEW} D bring further information. For the parameter sets presenting a well defined ballistic-like regime, most of the time the cells present  a well developed polarization and, hence, most of the points will present larger values of polarization. Cells will sporadically present small polarization when they are  changing the direction of their lamellipodia. On the other hand, in case the cells mostly present small values of polarization,  most of the time they are not in a ballistic-like regime. For $\phi=0.30$ and $\lambda_{F-actin}=150$,  most  points present small values of polarization. Observe that this happens in spite of a large value for $\phi$, when too much lamellipodium engulfing the cytoplasm  makes it more difficult to build a high polarization. Accordingly, this parameter set also shows a small value for the maximum exponent of the MSD curve, a short plateau at the $\langle u \rangle_{\delta}$ $versus$ $\delta$ plot, and a very small velocity-polarization correlation index.  In  Supplementary Materials Online Fig. S7  presents the equivalent of Fig. \ref{fig:Fig6NEW} for for $R_{cell}=10$ and $20$ and Figs. S8, S9, and S10 present the clouds of points, similar to Fig. \ref{fig:Fig6NEW} B, for all sets of parameters. Also, Figs. S11, S12, and S13  show the histograms for $\langle u(\Delta  \uptau_{opt}^{theor}) \rangle_{{\Pi}_{\beta} }  $ for all 4 polarization definitions and all different simulation sets, further detailing the information summarized in Fig. \ref{fig:Fig6NEW}. 

Now we turn to the direction of  cell velocity as compared to the polarization  $\vec{\Pi}_{CN-N}$. 
\begin{figure}[h]    
\centering
\includegraphics[width=1\textwidth]{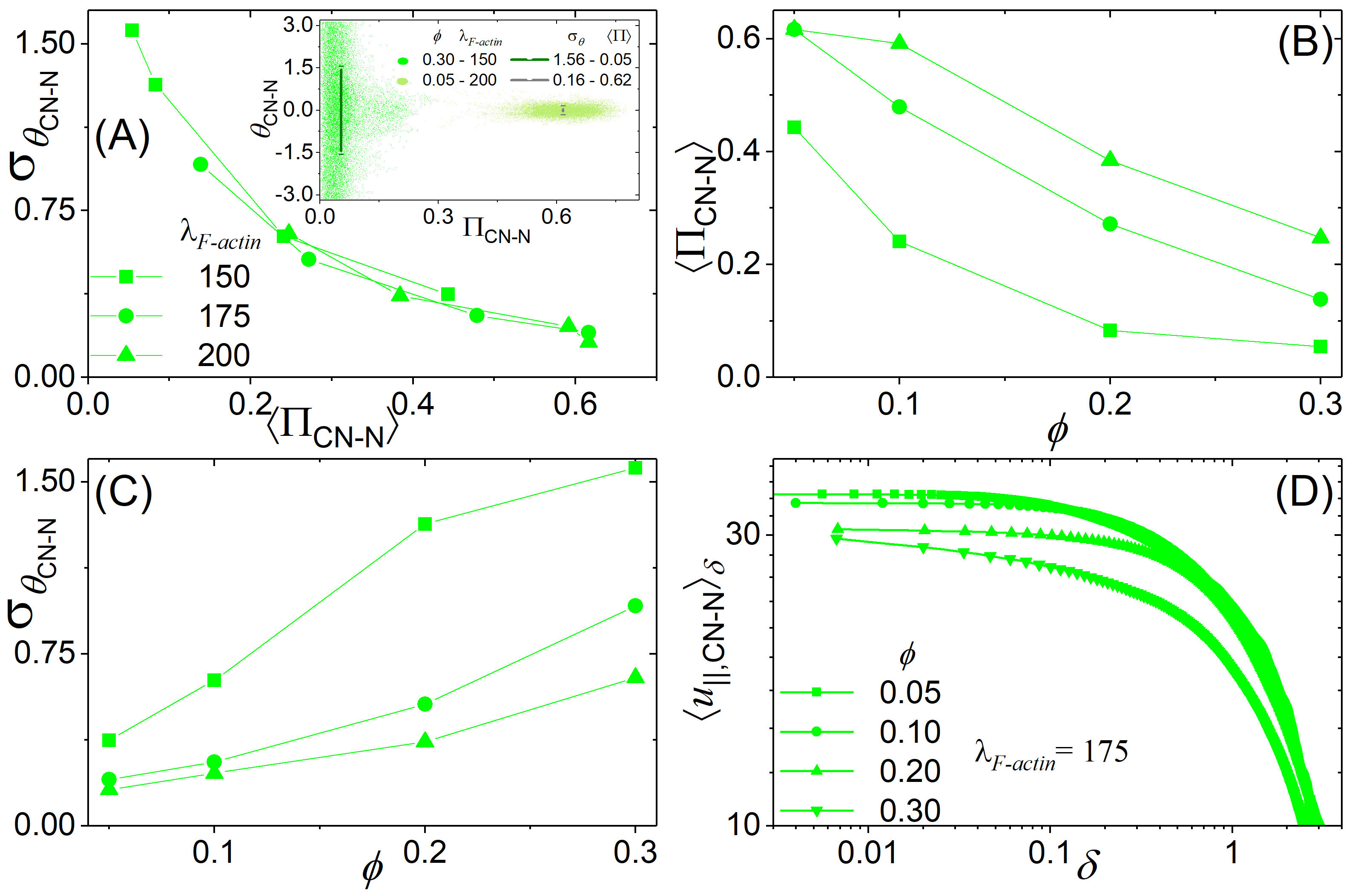}
\caption{ \label{fig:Fig7NEW} {\bf Performance of $\vec{\Pi}_{CN-N}$ as a predictor for cell displacement direction.} (A) The angular standard deviation $\upsigma_{\theta_{\mbox{\tiny CN-N}}}$  $versus$ average polarization modulus $\left <\Pi_{\mbox{\tiny CN-N}}(\uptau)\right >$   in units of cell radius for all parameter sets with $R_{cell} $=15. The displacements for the calculating $\theta_{\mbox{\tiny CN-N}}$ considered  $\delta=\Delta \uptau_{opt}^{theor}$. The inset presents individual measurements considering all points of the stationary trajectory, for 10 simulation runs. (B) Average polarization modulus $\langle {\Pi}_{CN-N}\rangle$  as a function of $\phi$. (C) $\upsigma_{\theta_{\mbox{\tiny CN-N}}}$ $versus$ $\phi$. (D) Plot of the average mean speed $\langle u_{\parallel,{\mbox{\tiny $CN-N$}}} \rangle_{\delta}$ measured parallel to $\vec{\Pi}_{CN-N}$ as a function of $\delta$.  Equivalent panels for $R_{cell} = 10$ and $20$ are shown in Figs. S14 and S15 in Supplementary Material Online.}
\end{figure}

Figure \ref{fig:Fig7NEW} A plots the standard deviation $\sigma_{\theta_{CN-N}}$ for the angle $\theta_{CN-N}$ between the mean velocity calculated using $\Delta\uptau_{opt}^{theor}$ and the cell polarization at the begining of $\Delta\uptau_{opt}^{theor}$ as a function of the average polarization modulus  $\langle \Pi_{CN-N} \rangle$ for each parameter set with $R_{cell}=15$. As expected,  $\sigma_{\theta_{CN-N}}$ decreases as $\langle \Pi_{CN-N} \rangle$ increases: more polarized cells show less deviation of their mean velocity from the polarization axis. The inset in this figure present the clouds for two sets of simulations: for both sets the mean value for $\theta_{CN-N}$ is zero (since the deviations from the polarization axis are symmetrical), but  the spread of the deviation angles are reduced for highly polarized cells. 

Figure~\ref{fig:Fig7NEW} B shows that the average polarization modulus decreases with $\phi$, reinforcing the conclusions drawn from Fig. \ref{fig:Fig6NEW}: a large fraction of
lamellipodium may not support a polarized front that would maintain a large polarization. To illustrate this point, we have produced two animations, SM1-15\_0.05\_200.mp4 and SM2-15\_0.30\_150.mp4, for the the two parameter sets considered in the inset of panel A, available in Supplementary Materials Online. Accordingly, the dispersion in direction increases with $\phi$, as shown in Fig.~\ref{fig:Fig7NEW} C. Finally, Fig.~\ref{fig:Fig7NEW} D shows that the average mean speed $\langle u_{\parallel,{\mbox{\tiny $CN-N$}}} \rangle_{\delta}$ measured parallel to $\vec{\Pi}_{CN-N}$ converges as $\delta \rightarrow 0$ and, hence, this component is well defined. In  Supplementary Materials Online Figs. S14 and S15  present the equivalent of Fig. \ref{fig:Fig7NEW} for for $R_{cell}=10$ and $20$ and Figs. S16, S17, and S18 present the clouds of points of the inset in Fig. \ref{fig:Fig7NEW} A for all sets of parameters. 

Together, Figs. \ref{Fig:Fig3NEW} to \ref{fig:Fig7NEW}  show that it is possible to define a polarization vector for the cells, and that it may serve as a proxy for cell displacement, provided the cells are at least minimally polarized.  They also show that the most adequate polarization definitions are those that do not consider the lamellipodium center of mass. Regarding $\beta= CN-N$ or $C-N$, for all parameter sets there is a clear  
correlation between the size of amplitude of polarization and both mean speed (calculated using $\delta=\Delta \uptau_{opt}^{theor}$) and polarization direction. We remark, however, that when the cells are in a poorly migrating phenotyped, as parameterized by $\phi$, persistent motion is rare and the polarization is small. In these cases, the definition of a polarization as a proxy for cell displacement is meaningless.

The correlation of polarization with cell displacement depends on  cell displacement being measured for a time interval equal to $\delta=\Delta\uptau_{opt}^{theor}$. In turn, the definition of $\Delta\uptau_{opt}^{theor}$ depends on assuming that cells kinematics may be characterized by a MSD curve given by the modified F\"urth equation, Eq. \ref{Eq:ModFurth} \cite{Fortuna2019}. While  F\"urth equation represents the stationary  solution of an isotropic Ornstein-Uhlenbeck problem for the cell velocity, the modified F\"urth equation is the stationary solution for an anisotropic Ornstein-Uhlenbeck problem \cite{deAlmeida2020}. In this model, the cell's instantaneous vector velocity is an ill-defined quantity but the speed in the direction of the polarization is well-defined and follows a one dimensional Langevin problem. 
As shown by the divergence of speed as $\delta \rightarrow 0$ in Fig. \ref{Fig:Fig3NEW} B, the present simulations are  able to reach the ill-defined velocity regime.
Since here we have defined a polarization vector, we may now verify whether the cell's  velocity in the direction of polarization converges or not as the time interval used to calculate the displacement decreases. 
The speed in the direction of the polarization is defined as
\begin{equation}
u_{\parallel,\beta}(\uptau,\delta) =  \vec{\Pi}_\beta(\uptau)\cdot\frac{\vec {\rho}(\uptau+\delta)-\vec {\rho}(\uptau)}{\delta}   \,\, , \,\,\,\beta = CN-N, L-CN, L-N, C-N  \, ,
\end{equation} 
and its average, $\langle u_{\parallel,\beta} \rangle_{\delta}$ for a given value of $\delta$, is shown in 
Fig. \ref{fig:Fig7NEW} D for $\beta = CN-N$. We may conclude that the speed in the direction of the polarization is a well defined quantity. Figs. S19, S20, and S21 in Supplementary Materials Online
show equivalent plots for all sets of parameters. 

\section{Discussion and Conclusions}

From the results we presented above, 
we reach the following conclusions.

First, it is possible to define a polarization vector on the basis of its correlation with cell displacement. Although we have considered simulations, we propose cell polarization definitions that are readily transposed to experiments, as the distance between the centers of nucleus and cytoplasm. For our simulations, polarization definitions that do not consider the position or shape of lamellipodia showed a better performance as cell displacement predictors. In migrating cells, lamellipodium position determines the localization of the nucleus behind the cell's  geometric center \cite{Gundersen2018}. Consequently  both nucleus and lamellipodium positions carry information on the direction of the movement. Nuclear positions, however, are more persistent and show smaller fluctuations.

Second, the assessment of the performance for polarization definition as a proxy for cell displacement direction depends on the time interval used to measure displacement.  The determination of the ideal time interval depends on the fitting of MSD curve using the modified F\"urth equation \cite{Fortuna2019,Thomas2019}.  The optimal time interval for obtaining cell displacement are shown in Fig. \ref{Fig:Fig5NEW} for all values of $S$, provided the simulated or experimental MSD curve is  well fitted by the modified F\"urth equation. A detailed procedure for fitting experimental MSD curve is provided in \cite{Thomas2019} and its associated supplementary materials, where the fitting was performed for 12 different experimental setups from 5 different laboratories. A second, faster but less precise method  is  1) Plot the mean square displacement curve (in a log-log plot)  and then 2) find the time interval that yields the maximum exponent in that curve. That would give an estimate for the optimal time interval in laboratory units to correlate polarization and mean speed. Either way, to obtain the polarization, it is necessary to follow the cell geometrical center of mass (without lamellipodium) and the cell nucleus.

Third, the performance of  the polarization vector is quantified by Figs. \ref{fig:Fig6NEW} and \ref{fig:Fig7NEW}.  The polarization vector is  a good predictor for cell displacement provided  it is not small. For cells that spend most of the time polarized, and hence have a migrating phenotype, our definitions apply well to the simulations. That must be  verified  in 
experiments. 

Fourth, the simulations verified that velocity is well defined when measured parallel to the polarization direction, that is, the average mean speed in the polarization direction, $\langle u_{\parallel, CN-N}\rangle_{\delta}$,  converges to a defined value as $ \delta \rightarrow 0$, in contrast to the average mean speed, $\langle u\rangle_{\delta}$, that diverges in the same limit, as shown in Fig. \ref{Fig:Fig3NEW}B. This behavior is in agreement with the theoretical predictions of the Anisotropic Ornstein-Uhlenbeck model \cite{deAlmeida2020}. In the direction perpendicular  to the polarization axis, the velocity is ill-defined, diverging when the time interval used for estimating displacement approaches zero. This must be verified in experiments. 
    
The  polarization definition and measurement that we propose can be applied to  migration experiments of single cells on flat substrates. Migration is an important feature in characterizing cell phenotypes as, for example, in cancer where tumor malignancy correlates with cell migration capabilities, or where the tumor cell diversity may indicate different disease outcomes. In these and many other instances, a robust and reproducible protocol for characterizing migrating cells is a first, necessary step. A second step is to obtain robust distributions of quantities (properly measured), such that phenotypic classes may be proposed, there is the need of obtaining  not only the value of quantities as $S$, $D$, $P$, or the mean speed $\Delta\uptau_{opt}^{theor}$, but also their distributions in a given population. This is a work in progress. 

The extension of these definitions and measurement procedures to collectively migrating cells may also contribute to the field. This is another work in progress and will be presented elsewhere.

\section*{Acknowledgemnts}
We thank Dr. H\'el\`ene Delano\"e-Ayari, of Universit\'e Claude Bernard - Lyon 1, and Dr Andrew Callan-Jones, from Universit\'e Paris Diderot, for carefully reading and providing suggestions for the manuscript.
This work has been performed under the CAPES-COFECUB agreement 88887/190134.
We acknowledge partial support from Brazilian Agency Conselho Nacional de Desenvolvimento Científico e Tecnológico - CNPq.

\bibliographystyle{unsrt}
\bibliography{main} 

\end{document}